\documentclass[preprint2,twoside]{hwo}

\usepackage{graphicx}

\input{hwo.h}

\begin{document}

\title{\textbf{\LARGE Exoplanet Atmospheric Escape Observations with the Habitable Worlds Observatory}}
\author {\textbf{\large Leonardo A. Dos Santos,$^{1,2}$ Eric D. Lopez,$^{3, 4}$}}
\affil{$^1$\small\it Space Telescope Science Institute, 3700 San Martin Drive, Baltimore, MD 21218, USA; \email{ldsantos@stsci.edu}}
\affil{$^2$\small\it Department of Physics and Astronomy, Johns Hopkins University, 3400 N. Charles Street, Baltimore, MD 21218, USA}
\affil{$^3$\small\it NASA Goddard Space Flight Center, 8800 Greenbelt Rd, Greenbelt, MD 20771, USA}
\affil{$^4$\small\it GSFC Sellers Exoplanet Environments Collaboration, NASA GSFC, Greenbelt, MD 20771, USA}

\author{\small{\bf Contributing Authors:} Luca Fossati (Space Research Institute, Austrian Academy of Sciences), Antonio Garc\'ia Mu\~noz (CEA Paris-Saclay), Shingo Kameda (Rikkyo University), Munazza K. Alam (Space Telescope Science Institute), Keighley Rockcliffe (CRESST/UMBC, NASA/GSFC), Seth Redfield (Wesleyan University), Yuichi Ito (National Astronomical Observatory of Japan), Joshua Lothringer (Space Telescope Science Institute), Shreyas Vissapragada (Carnegie Observatories), Hannah R. Wakeford (University of Bristol), Apurva V. Oza (California Institute of Technology), Girish M. Duvvuri (Vanderbilt University), Raissa Estrela (Jet Propulsion Laboratory, California Institute of Technology), Ryoya Sakata (The University of Tokyo), Chuanfei Dong (Boston University), Ziyu Huang (Boston University)}


\author{\footnotesize{\bf Endorsed by:} Katherine Bennett (Johns Hopkins University), Abby Boehm (Cornell University),  Aarynn Carter (Space Telescope Science Institute), Miguel Chavez Dagostino (National Institute of Astrophysics, Optics and Electronics), Caleb Harada (UC Berkeley), James Kirk (Imperial College London), Adam Langeveld (Johns Hopkins University), Eunjeong Lee (EisKosmos (CROASAEN), Inc.), Drew Miles (California Institute of Technology), Faraz Nasir Saleem (Egypt Space Agency), Gaetano Scandariato (INAF), Jessica Spake (Carnegie Observatories), Christopher Stark (NASA/GSFC), Antoine Strugarek (CEA Paris-Saclay, DAp-AIM), Johanna Teske (Carnegie Earth and Planets Laboratory), Daniel Valentine (University of Bristol), Austin Ware (Arizona State University), Peter Wheatley (University of Warwick)
}



\begin{abstract}
  The Decadal Survey on Astronomy and Astrophysics 2020 highlights the importance of advancing research focused on discovering and characterizing habitable worlds. In line with this priority, our goal is to investigate how planetary systems evolve through atmospheric escape and to develop methods for identifying potentially Earth-like planets. By leveraging the ultraviolet (UV) capabilities of the Habitable Worlds Observatory, we can use transit spectroscopy to observe atmospheric escape in exoplanets and explore the processes that shape their evolution, assess the ability of small planets to retain their atmospheres, and search for signs of Earth-like atmospheres. To achieve this, we support the development of a UV spectrograph with moderate- to high-resolution capabilities for point-source observations, coverage of key spectral features in the 1000–3000~\AA\ range, and detectors that can register high count rates reliably. \textit{This article is an adaptation of a science case document developed for the Characterizing Exoplanets Steering Committee within HWO's Solar Systems in Context Working Group.}
  \\
  \\
\end{abstract}

\vspace{2cm}

\section{Science Goal}

The fundamental question we aim to address in this HWO science case is: How efficiently can planets retain their atmospheres and thus be habitable?

Most transiting exoplanets orbit very closely to their host stars, subjecting them to irradiation levels that are 10 to 1000 times that experienced by the Earth, which can lead to significant atmospheric loss due to evaporation. This effect has an outsized impact on whether these planets have conditions amenable to life, as the presence of an atmosphere is seen as an important ingredient for habitability. Unlike solar system planets, for which intense evaporation is no longer detectable, exoplanets represent our best opportunity to see the phenomenon occurring now. To answer the aforementioned questions, we need to develop a survey to observe atmospheric escape in exoplanets within a range of physical parameters -- including those similar to Earth -- and develop accurate models to interpret these observations.

\subsection{Topics Related to the Astro2020:}
\begin{itemize}
    \item \textit{The Distribution and Nature of Sub-Neptune Planets:} Understanding the role of atmospheric escape in  the evolution of sub-Neptunes.
    \item \textit{Exoplanet Characterization and Solar System Synergy:} Investigating the physical properties of exoplanet atmospheres and how they interact with the irradiation environment.
    \item \textit{The Dawn of Exoplanet Astrobiology: The Search for Habitable Environments and Life:} Enabling the search for Earth-like exospheres in transiting exoplanets.
\end{itemize}

In particular, this Science Case is relevant to the following Key Science Questions and Discovery Areas of the Astro2020 Decadal Survey Report:

\begin{itemize}
    \item \textbf{E-Q2.} What are the properties of individual planets, and which processes lead to planetary diversity?
    \begin{itemize}
        \item \textbf{E-Q2b.} How does a planet’s interior structure and composition connect to its surface and atmosphere?
        \item \textbf{E-Q2c.} What fundamental planetary parameters and processes determine the complexity of planetary atmospheres?
        \item \textbf{E-Q2d.} How does a planet’s interaction with its host star and planetary system influence its atmospheric properties over all time scales?
        \end{itemize}
    \item \textbf{E-Q3.} How do habitable environments arise and evolve within the context of their planetary systems?
    \begin{itemize}
        \item \textbf{E-Q3a.} How are potentially habitable environments formed?
        \item \textbf{E-Q3b.} What processes influence the habitability of environments?
        \item \textbf{E-Q3c.} What is the range of potentially habitable environments around different types of stars?
        \item \textbf{E-Q3d.} What are the key observable characteristics of habitable planets?
    \end{itemize}
\end{itemize}

\subsection{Relevance for Other Broad Scientific Areas in the Astro2020 Decadal Survey:}
\begin{itemize}
    \item Panel on the Interstellar Medium and Star and Planet Formation (F): Synergy with studies of the interstellar medium in the ultraviolet.
    \item Panel on Optical and Infrared Observations from the Ground (K): Synergy with high-resolution near-infrared spectroscopy.
    \item Panel on Electromagnetic Observations from Space 1 (I): Synergy with current and future ultraviolet, optical and near-infrared space instruments.
    \item Panel on Electromagnetic Observations from Space 2 (J): Synergy with current and future X-rays and extreme-UV space instruments.
\end{itemize}

\section{Science Objective}

In this science case, we focus on the topic of atmospheric escape and evolution in exoplanets. Our science objective can be broken down into two goals, which we discuss below.

\subsection{Determine whether [transiting] rocky planets in habitable zones have exospheres similar to the modern Earth}

\begin{figure*}[ht!]
    \centering
    \includegraphics[width=0.75\textwidth]{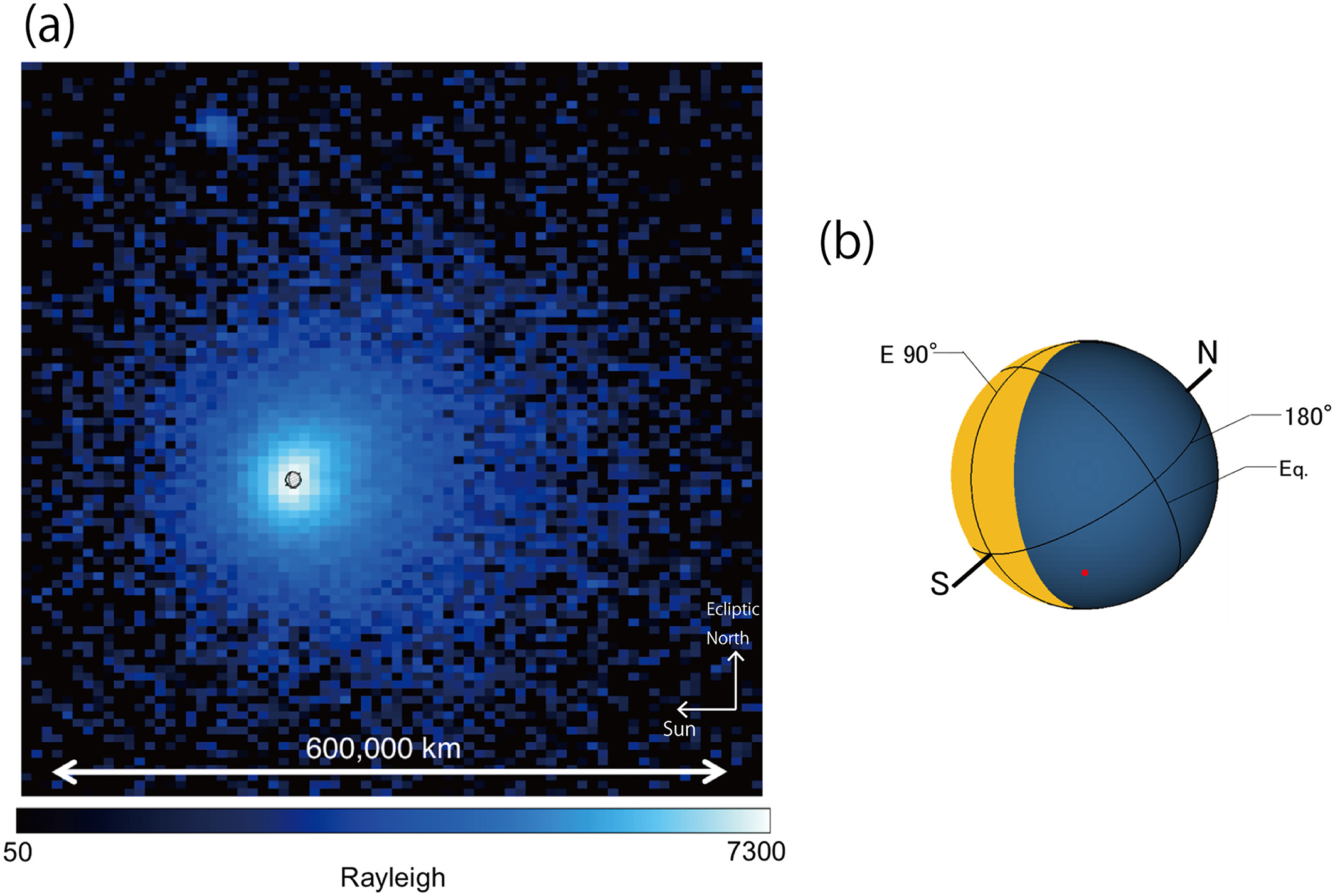}
    \caption{(a) The Earth’s extensive and H-rich exosphere scatters solar Lyman-$\alpha$ photons, producing the geocoronal airglow; (b) the geometry of Earth’s illumination by the Sun \citep[adapted from][]{Kameda2017}. Due to its size, such a large exosphere can produce a potentially detectable absorption of stellar Lyman-$\alpha$ photons when observed in transit \citep{Castro2018, DSantos2019a}.}
    \label{fig:fig1}
\end{figure*}

The outermost layer of a planet’s atmosphere is known as its exosphere, where the density of the gas is so low that collisions between particles are unlikely to happen yet the density is high enough that the gas is detectable at a select few spectral lines. The exosphere of the modern Earth is composed mostly of neutral atoms of hydrogen (H), oxygen (O) and nitrogen (N) \citep{Shizgal1996}. Most importantly, the Earth has a unique exosphere compared to other rocky planets in the Solar System, particularly in its extension (more than 40 Earth radii; see Figure \ref{fig:fig1}) and H-rich composition \citep{Kulikov2007}. The situation is very different for Venus and Mars because their atmospheres are either water-deficient or not sufficiently extended. Thus, detecting a similarly large and H-rich exosphere around a transiting exoplanet could provide compelling evidence to corroborate whether this planet is similar to modern Earth (i.e., it hosts a water reservoir) or not. Alternatively, it may provide insight into other alternative compositions, such as those proposed for water-rich planets.

The Earth’s exosphere is replenished with H atoms by atmospheric escape at a rate of approximately $10^3$~g\,s$^{-1}$. These H atoms are produced by the photodissociation of water molecules from the layers below the exosphere. Thus, the search for H-rich exospheres around transiting planets can be used to probe evolved oceans \citep{Jura2004}.

\citet{Kameda2017} calculated a model of the Earth’s exosphere based on observations of Earth’s geocoronal emission. This model was then used by \citet{DSantos2019a} to estimate the observable signal of the Earth’s exosphere as an exoplanet transiting M dwarfs within a range of physical parameters. They conclude that this signal cannot be detected with the \emph{Hubble Space Telescope} (\emph{HST}), but the next NASA flagship mission, by then known as LUVOIR, can achieve this feat within a reasonable observing time.

\subsection{Determine how efficiently hydrodynamic escape erodes H2-dominated primordial envelopes}

\begin{figure*}[ht!]
    \centering
    \includegraphics[width=0.95\textwidth]{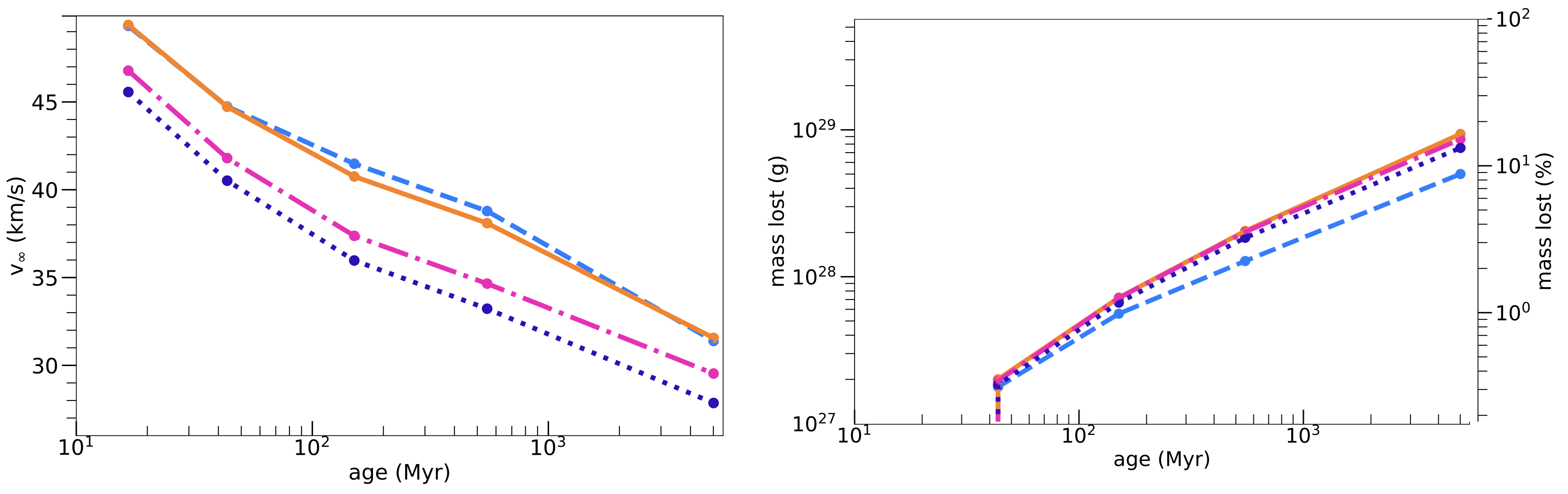}
    \caption{{\it Left panel}: Evolution of the mass-loss rate of a 0.3 MJup gas giant at a distance of 0.045 au from a K-type host star as a function of its age for four different atmospheric escape models \citep[see more details in][]{{Allan2024}}. {\it Right panel:} Total planetary mass lost as a function of age. The planet loses up to 10\% of its original mass over the course of 4 Gyr \citep[figure adapted from][]{Allan2024}.}
    \label{fig:fig2}
\end{figure*}

Atmospheric escape is a key factor shaping the evolution of planets \citep[e.g.,][see also Figure \ref{fig:fig2}]{Lammer2014, Owen2017, Kubyshkina2024}. It has a major impact on our understanding of planet formation and demographics. For example, accounting for mass loss for present-day planets makes it possible to extrapolate back in time their characteristics to derive their initial properties, e.g., mass of the accreted primordial H/He atmosphere \citep{Jin2014, Affolter2023}. However, despite being key to the atmospheric properties of planets, including potential habitability, the dependence of a planet’s atmospheric mass-loss on stellar (i.e., irradiating stellar UV flux and wind) and planetary (i.e., mass, radius, orbital separation) properties is far from understood \citep[e.g.,][]{Cockell2016,Lammer2018,Dong2018,Chin2024}. 

\begin{figure}[ht!]
    \centering
    \includegraphics[width=0.48\textwidth]{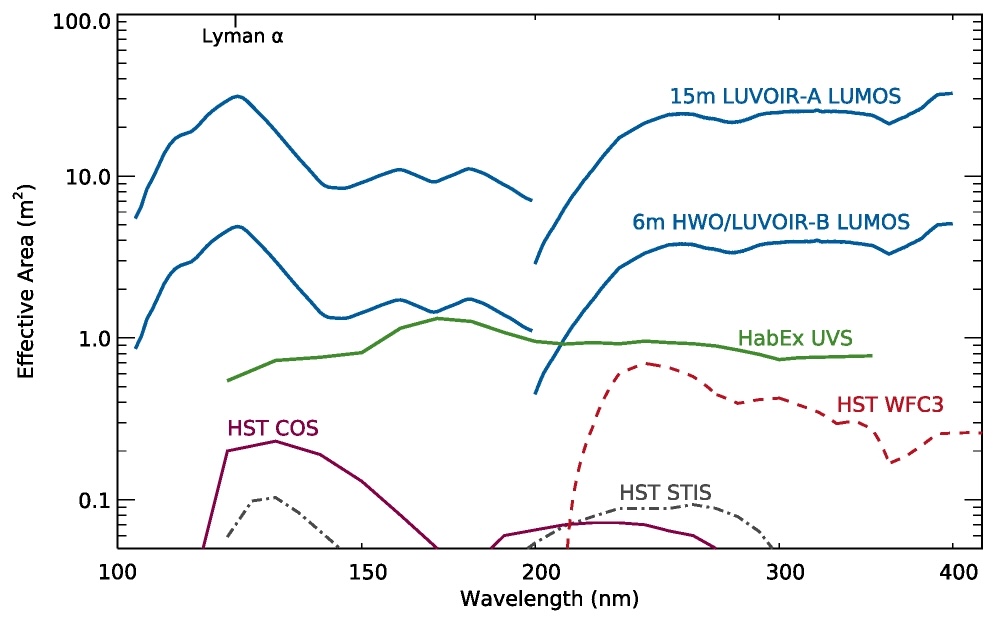}
    \caption{Effective area of different instruments capable of UV spectroscopy. According to its concept studies \citep{Bolcar2017a, Bolcar2017b, France2017}, the HWO/LUMOS instrument would have an effective area that is more than ten times better than the COS and STIS instruments on HST, assuming the more conservative telescope aperture size (LUVOIR-B).}
    \label{fig:fig3}
\end{figure}

To determine how efficiently hydrodynamic escape erodes primordial H/He envelopes in exoplanets, we aim to constrain atmospheric escape rates for exoplanets over a large range of masses and ages. UV transit spectroscopy has been one of the main pathways of measuring escape rates in short-period exoplanets \citep[see the review in][]{DSantos2023a}. This inference is obtained by fitting the transit light curve observations with models that include the relevant thermospheric and exospheric chemical components (e.g., H, C, O) and relevant physical processes (e.g., photoionization). Due to its predicted effective area in the UV (Figure \ref{fig:fig3}), HWO is uniquely poised to enhance our capabilities to observe photoevaporation in a large range of exoplanet masses.

\section{Physical Parameters}

This Science Case aims to leverage HWO's UV capabilities to measure the properties of atoms, ions and molecules in the upper atmospheres of transiting planets.

\subsection{Composition, ionization state, and size relative to the planet}

The Earth’s exosphere is composed mostly of neutral H and spans a geocentric distance of at least 38 Earth-radii \citep{Kameda2017}. The exospheric atoms remain neutral at a relatively long distance from the surface because their photoionization lifetime is approximately 20 days, on average \citep{Baliukin2019}. The ionization lifetime is dictated by the magnetic activity levels of the host star, since this process is dominated by extreme-UV irradiation.

\begin{figure}[ht!]
    \centering
    \includegraphics[width=0.45\textwidth]{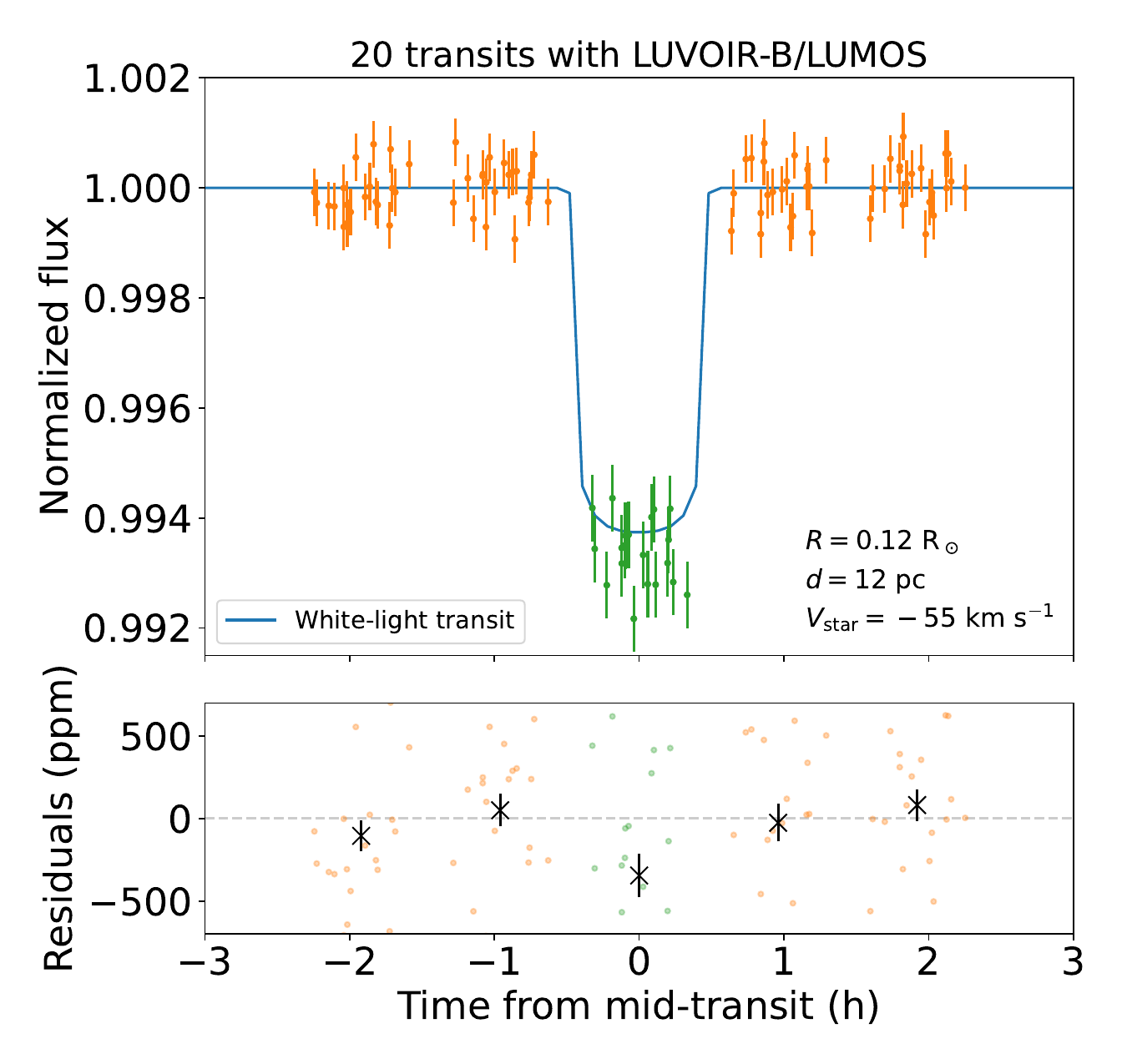}
    \caption{Simulated Lyman-$\alpha$ light curve of an Earth-like planet around TRAPPIST-1 planet using LUVOIR-B/LUMOS obtained by co-adding 20 transits. The planet produces a 500 ppm excess absorption compared to the white-light transit light curve, which is indicative of a large neutral-H cloud around the planet fed by atmospheric escape \citep[adapted from][]{DSantos2019a}.}
    \label{fig:fig4}
\end{figure}

In transiting exoplanets, we can measure their exospheric properties by observing their Lyman-$\alpha$ transit light curves. If the planet has a large neutral H cloud fed by atmospheric escape, it will imprint an excess absorption in the stellar Lyman-$\alpha$ line, whose depth is proportional to the size, ionization and density of the cloud. Previous attempts at observing this feature around potentially Earth-like exoplanets around TRAPPIST-1 with HST only yielded non-detections \citep{Bourrier2017a, Bourrier2017b, Berardo2025}. \citet{DSantos2019a} found that the Earth’s exosphere absorbs up to approximately 500 ppm of the stellar light in Lyman-$\alpha$, depending on the characteristics of the stellar host (see Figure \ref{fig:fig4}). If we aim to characterize the exospheres of Earth-like exoplanets using HWO, we should strive to achieve this level of precision in Lyman-$\alpha$ fluxes. 

The properties of the planet’s exosphere can, in turn, be measured by modeling the planet’s exosphere and comparing the theoretical transit light curve to the observed one in a Bayesian retrieval framework. Although there are, currently, theoretical models available to study a terrestrial planet's exosphere using a 3D Monte Carlo code \citep[e.g., see][]{Lee2021}, these are computationally costly. Thus, we will need to dedicate a substantial amount of work into creating the framework that will allow us to perform these inferences in a timely manner.

\subsection{Atmospheric escape rates and outflow compositions for 50 transiting exoplanets}

The atmospheric escape (or mass loss) rate of a planet is the main parameter that dictates their evolution, particularly for sub-Jovian mass planets. Currently, the most used technique to measure mass-loss rates is by observing the metastable helium (He) transmission spectrum \citep{Seager2000, Oklopcic2018} at high spectral resolution from the ground \citep{Allart2018, Salz2018, Nortmann2018}. The measurement is done by fitting a family of theoretical transmission spectra calculated with one-dimensional models to the observed spectrum, and retrieving the rates that best match the observation \citep{DSantos2022a}. The main limitation of this technique is that, at first glance, seems to favor planets orbiting K-type stars \citep{Oklopcic2019}. Thus, at present date, surveys for atmospheric escape in exoplanets have been biased towards these cases \citep[see, however,][]{Bennett2023}. Additionally, ground-based observations in H$\alpha$ transmission spectroscopy have been used as escape tracers \citep[e.g.,][]{Jensen2012, CBarris2019, Yan2021}, which more directly traces hot hydrogen thermospheres. The limitation of H$\alpha$ is that it has more predominantly been detected in ultra-hot Jupiters and it is highly sensitive to stellar activity.

\begin{figure*}[ht!]
    \centering
    \includegraphics[width=0.95\textwidth]{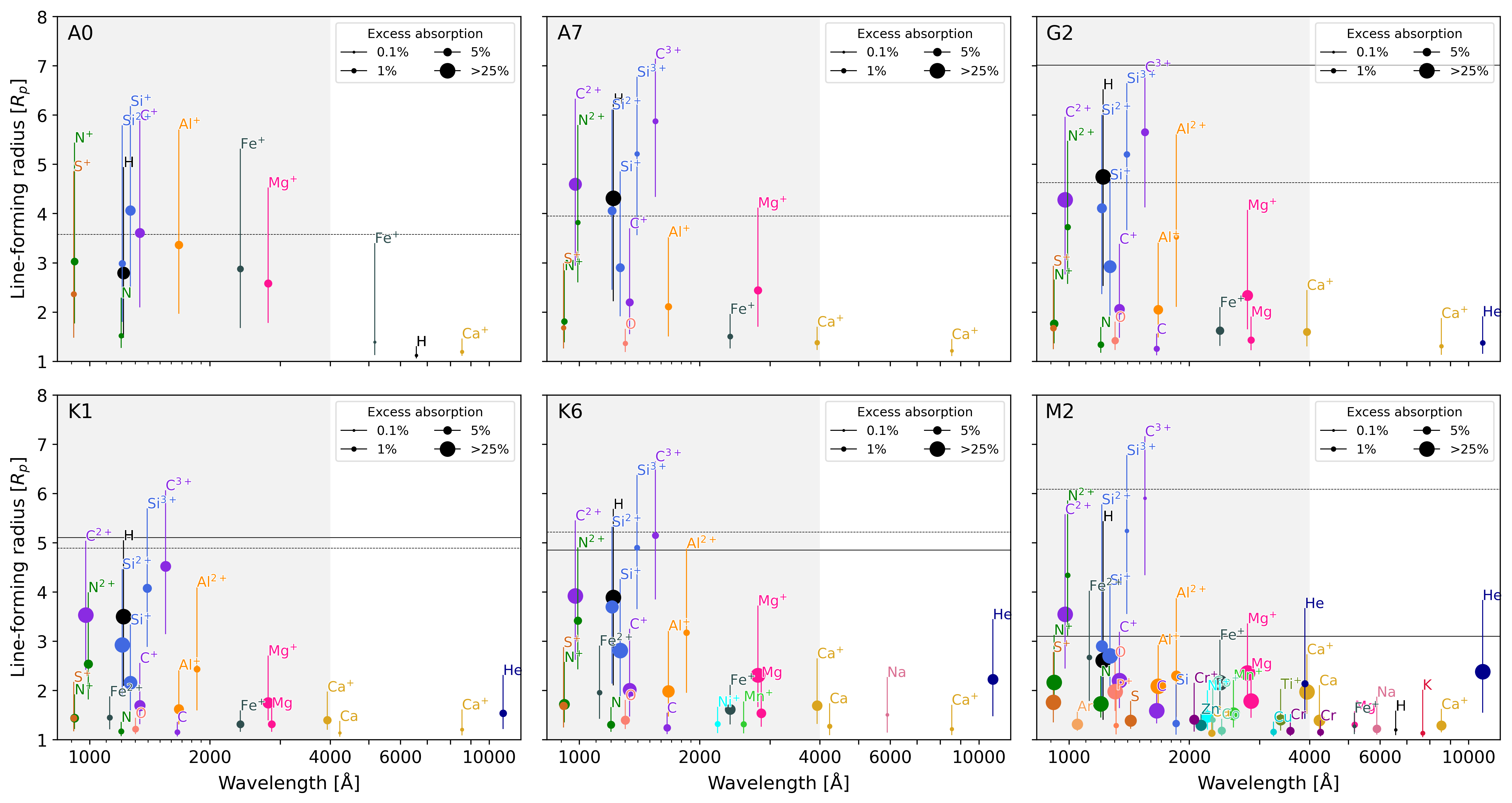}
    \caption{Spectral features detectable in transmission spectra of hot Jupiters as a function of wavelength and stellar host type \citep[adapted from][]{Linssen2023}. The wavelengths relevant for UV spectroscopy are those below 200 nm. The line-forming radius (y-axis) is a proxy for the altitude where the feature is located in relation to the planetary surface and the symbol sizes represent the depth of the excess absorption in transmission.}
    \label{fig:fig5}
\end{figure*}

Another technique that can also be used to infer mass-loss rates is by observing UV transmission spectra. At these wavelengths, we have access to a large selection of spectral lines (as opposed to the metastable He, which is a triplet). An investigation by \citet{Linssen2023} found that many of these can be used as tracers for atmospheric escape in transiting exoplanets (see Figure \ref{fig:fig5}). 

The theoretical interpretation of metastable He and H$\alpha$ observations is highly non-trivial, since they require detailed knowledge of many microphysical processes (such as excitation/de-excitation rates and Lyman-$\alpha$ cooling) to retrieve accurate mass-loss rates. On the other hand, in the UV, these tracers are predominantly resonance lines, so their interpretation is more straightforward compared to ground-based observables.

Furthermore, one important limitation of using only one observable feature (e.g., the near-IR metastable He triplet) to study atmospheric escape is that there is a wide range of models and parameters that can explain the observations, causing degeneracies. A well known example is the degeneracy between the mass-loss rate and the outflow temperature when assuming the isothermal Parker wind formulation \citep{Vissapragada2022} to fit metastable He transmission spectra. By using many lines, we can break these model degeneracies, which highlights the importance of enabling high-resolution spectroscopy in the UV with HWO. Multiple tracers also allow us to better constrain the level of activity contamination.

The interplay between a multi-species study of atmospheric escape in the UV has already been illustrated for the transiting sub-Neptune $\pi$~Men~c, which was observed with HST/STIS \citep{GMunoz2020, GMunoz2021}. In this study, the authors provide compelling evidence that the planet likely has a high-mean molecular weight atmosphere, a result that would otherwise been inconclusive if the planet was observed in the optical or near-IR, where the atmospheric metallicity is degenerate with the presence of clouds \citep[e.g.,][]{Kreidberg2014}. Finally, observations of escaping metals enable us to probe magnetic fields in exoplanets, as seen for HAT-P-11b with HST \citep{BJaffel2022}.

\subsection{Observational State of the Art}

The state of the art parameters correspond to roughly what can be observed for hot Jupiters and Neptunes today with HST. For the incremental progress, the parameters listed in Table \ref{tab:performance} correspond to roughly a sub-Neptune with a H/He envelope. Substantial progress represents the expected parameters for a young Earth with some primordial H/He. Major progress corresponds to an Earth-like planet.

\begin{table*}[ht!]
    \centering
    \caption[Physical Parameters]{Exospheric physical parameter benchmarks for HWO configurations.}
    \label{tab:performance}
    \begin{tabular}{lcccc}
        \noalign{\smallskip}
        \hline
        \noalign{\smallskip}
        {Physical Parameter} & {State of the Art} & {Incremental Progress} & {Substantial Progress} & {Major Progress} \\
        \noalign{\smallskip}
        \hline
        \noalign{\smallskip}
        Exospheric radius & $\sim 90$~R$_\oplus$ & $\sim 70$~R$_\oplus$ & $\sim 50$~R$_\oplus$ & $40$~R$_\oplus$ \\
        Exospheric density & $\sim 10^9$~cm$^{-3}$ & $\sim 10^8$~cm$^{-3}$ & $\sim 10^7$~cm$^{-3}$ & $10^5$~cm$^{-3}$ \\
        Exobase temperature & $T > 7000$~K & $T> 7000$~K & $T > 3000$~K & $1000$~K \\
        Atmospheric escape rate & $\dot{m} > 10^9$~g\,s$^{-1}$ & $10^7$~g\,s$^{-1}$ & $10^5$~g\,s$^{-1}$ & $10^3$~g\,s$^{-1}$ \\
        \noalign{\smallskip}
        \hline
    \end{tabular}
\end{table*}

\section{Description of Observations}

We propose two approaches to leverage HWO's UV capabilities for studies of atmospheric escape and exoplanet evolution: a ``deep-field"-like program targeting one or more high-profile exoplanets and a broad survey targeting a wide range of exoplanets.

\subsection{A ``deep-field"-like campaign to detect an Earth-like exosphere around a rocky planet}

Following the recommendations of \citet{DSantos2019a}, we propose to execute a dedicated campaign to detect an Earth-like exosphere using HWO and its UV spectrograph (equivalent to LUMOS). The authors estimated that the excess absorption caused by a cloud of neutral H around an Earth-like exoplanet around an M dwarf is approximately 500~ppm in relation to the opaque-disk transit depth of the planet. While a 500~ppm absorption is well-within the capabilities of near-IR detectors, such as those in JWST \citep{JTEC2023}, it is paramount that the UV spectrograph of HWO has access to this level of spectrophotometric precision when co-adding 20 or less transits.

Special care should be taken when selecting a target for this deep-field campaign. Besides the requirements that the planet should be transiting, rocky and within the habitability zone (HZ), the star needs to be nearby enough and have a high-enough radial velocity to minimize interstellar medium (ISM) absorption of the Lyman-$\alpha$ line. With a distance of about 12.5~pc, a radial velocity of -52~km\,s$^{-1}$, and more than one planet within its HZ, TRAPPIST-1 is currently the best target for this deep campaign \citep{Gillon2017}. However, depending on how many other amenable candidates are discovered until the launch of HWO, more planets could be included in this campaign.

Points for further research in preparation for this campaign:

\begin{itemize}
    \item Identifying other optimal target systems, including those around stars of different spectral types \citep[see, e.g.,][]{CAguirre2023}.
    \item Stellar activity contamination and timescales in the Lyman-$\alpha$ line, particularly for M and K dwarfs.
    \item Simulations of escaping O and N\footnote{The detectability of N\,{\sc i} in the upper atmosphere of an Earth-like exoplanet with LUVOIR-A has already been investigated in \citet{Young2020}. However, further studies in the case of hotter thermospheres (around low-mass stars) and detectability of N\,{\sc ii} at 108~nm are still warranted.} in Earth-like exoplanets to determine whether these signals are detectable.
    \item Simulations of escaping C in Venus-like exoplanets.
    \item Atomic signatures in transmission spectra of planets similar to the young Earth.
    \item Although HWO is being designed to detect habitable worlds through direct methods, it is still unclear whether it is technically feasible to observe atmospheric escape using direct spectroscopy \citep[e.g.,][]{Rosener2025}, so more research on this topic is encouraged.
\end{itemize}

\subsection{A broad survey of atmospheric escape of metals in a large sample of transiting exoplanets}

\begin{figure}[ht!]
    \centering
    \includegraphics[width=0.45\textwidth]{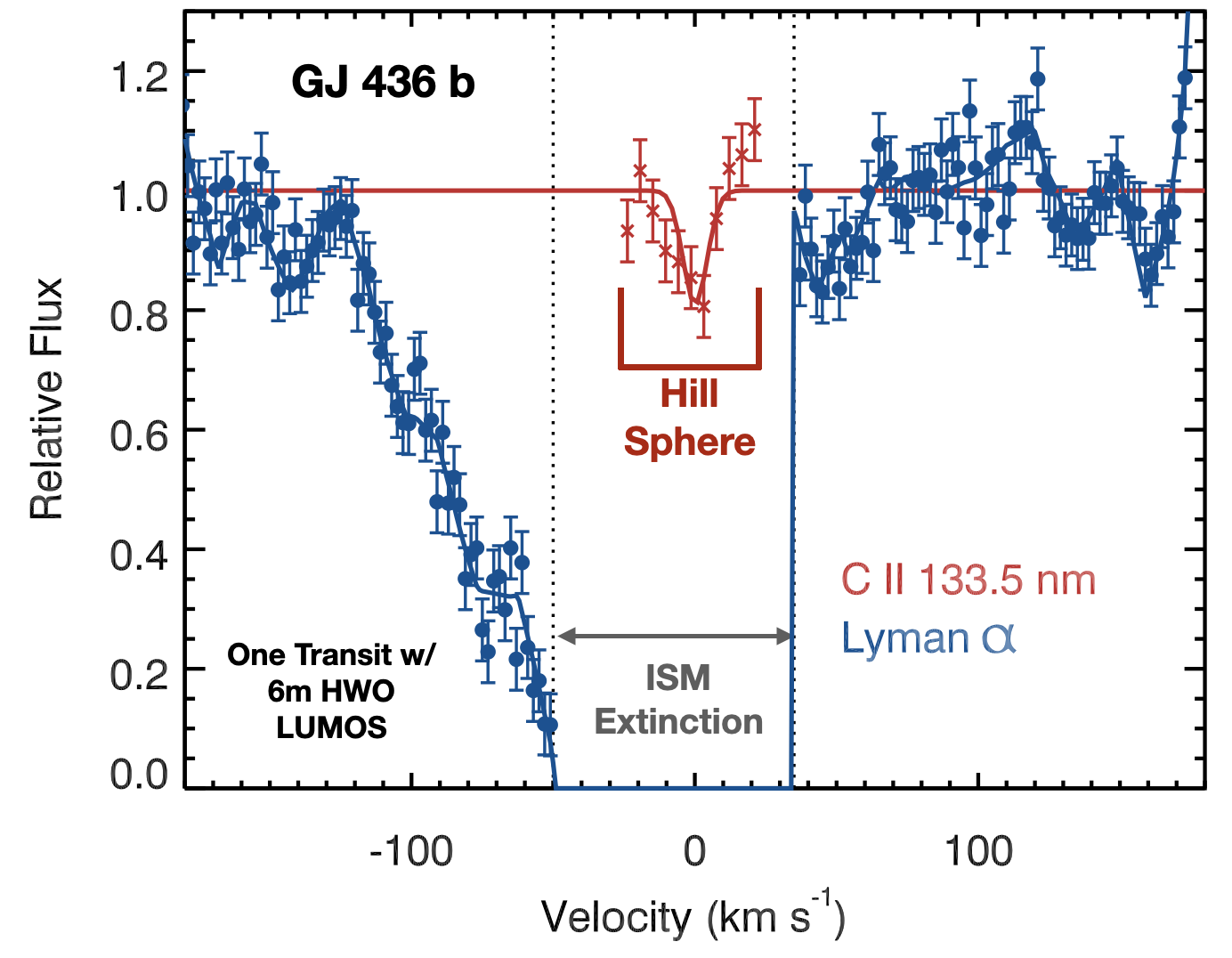}
    \caption{Simulated UV transmission spectrum of the Neptune-sized exoplanet GJ~436~b in the Lyman-$\alpha$ line (which traces escaping hydrogen; blue) and singly-ionized carbon (red). These spectra are shown in Doppler velocity space (x-axis) instead of wavelength space, and null velocity corresponds to the rest wavelength of the spectral line in the stellar rest frame. This simulation assumes one transit with HWO, with a 6-m telescope aperture and the original characteristics of the LUMOS spectrograph. Figure is adapted from LUVOIR STDT; E. Lopez (NASA/GSFC) \& K. France (Univ. Colorado).}
    \label{fig:fig6}
\end{figure}

In contrast to the deep campaign explained above, we propose a broad survey of atmospheric escape signatures in approximately 50 transiting exoplanets within a range of planetary properties, varying from hot Jupiters to cool sub-Neptunes. Building on past efforts that studied transits in Lyman-$\alpha$ (121~nm) and C\,{\sc ii} (133~nm), we simulated how these two signatures would present themselves in one transit with HWO (see the example of the warm Neptune GJ 436b in Figure \ref{fig:fig6}). These represent only two of the signals that can be detected but, as seen in \citet{Linssen2023}, there is a wide range of other spectral tracers available in the UV that should be included in the wavelength access of HWO and its UV instrument.

\begin{figure*}[ht!]
    \centering
    \includegraphics[width=0.95\textwidth]{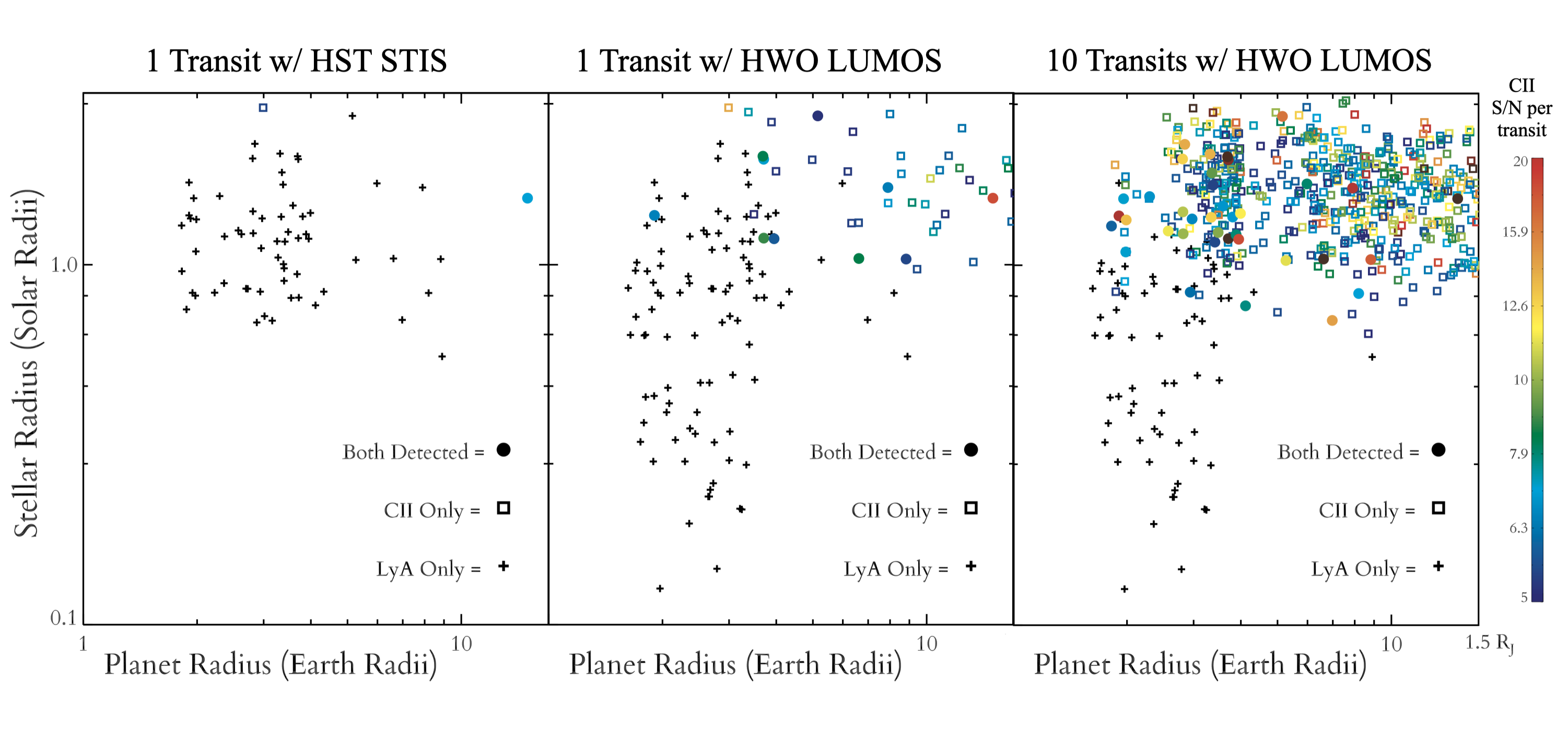}
    \caption{Simulated population of exoplanets that can have significant detections of escaping hydrogen and/or ionized carbon if observed with HST/STIS and HWO/LUMOS in a given number of transits. By co-adding several transits with HWO it will be possible to fully characterize escape from dozens of exoplanets across parameter space.}
    \label{fig:fig7}
\end{figure*}

Figure \ref{fig:fig7} shows our preliminary estimate of what might be possible with such a UV transit survey. Using a simulated catalog of transiting planets found by TESS \citep{Barclay2018}, we model the expected S/N in Lyman-$\alpha$ and the 133~nm C\,{\sc ii} doublet for planets of different sizes and host star types. On the left we highlight what is currently possible with HST: Lyman-$\alpha$ detections for a few dozen nearby planets and low S/N C\,{\sc ii} detections for the very best hot Jupiters. In contrast, by co-adding multiple transits with HWO we should be able to detect C\,{\sc ii}, and many other metal lines, at very high S/N for dozens of planets all the way down to sub-Neptune sizes. 

Such a survey is not currently possible with HST because of its significantly lower UV throughput compared to HWO. Although some spectral tracers are available in the optical for ground-based observatories (see Figure \ref{fig:fig5}), they are much less numerous than those in the UV.

Another important requirement for this campaign is related to how resilient the UV detector is to high count rates (i.e. dynamic range). With HST, both the STIS and COS UV detectors have stringent bright-object observing limits due to their design and loss of sensitivity over time following consecutive observations. With the larger mirror of HWO, we would like to observe bright stars to detect the faint signals of exoplanet atmospheres in the UV, thus the detectors need to be more resilient to high count rates than those currently installed on HST. While the use of filters could be a remedy to this issue, it would also decrease the signal-to-noise ratios of the observations since they are photon-limited, thus filters should not be the main solution.

A high spectral resolution (similar to HST/COS or better) is required for these observations, since they are based on a line-by-line analysis \citep[see, e.g.,][]{DSantos2019b}. While resolutions as high as STIS in the {\it \'echelle} mode are not strictly required for this science case, resolutions similar to those achieved in ground-based spectrographs ($R \sim 45,000$) would enable us to distinguish stellar activity effects, detect the planetary Doppler motion across the star \citep{Cegla2016}, and better infer the properties of the ISM \citep{Redfield2004}.

Points for further research in preparation for this campaign:

\begin{itemize}
    \item Determine more exact spectral resolution constraints for transiting exoplanet observations. A spectral resolution similar to HST/COS is, generally, acceptable at the moment, but research on the impact of planetary Doppler motion is needed to determine which specific observations require a better resolution, and quantify the need.
    \item Identify sources of opacity in exoplanet atmospheres at UV wavelengths.
    \item Combine models of lower atmospheres and upper atmospheres to better interpret future observations.
\end{itemize}

We present the observation requirements for the campaigns described above in Table \ref{tab:obsreq}, which correspond to the both the FUV and NUV detectors. The state of the art corresponds to current capabilities with HST and the STIS and COS spectrographs. A more detailed discussion of the proposed observational campaigns and the technical requirements of HWO's UV spectrograph are presented in an upcoming manuscript submitted to the Journal of Astronomical Telescopes, Instruments and Systems (Dos Santos \& Lopez, under review).

\begin{table*}[ht!]
    \centering
     \caption[Observation Requirements]{Requirements for detecting atmospheric escape in exoplanets with HWO.}
    \label{tab:obsreq}
    \begin{tabular}{lcccc}
        \noalign{\smallskip}
        \hline
        \noalign{\smallskip}
        {Obs. Requirement} & {State of the Art} & {Incremental Progress} & {Substantial Progress} & {Major Progress} \\
        \noalign{\smallskip}
        \hline
        \noalign{\smallskip}
        Effective area & $680$~cm$^2$ & $6\,800$~cm$^2$ & $73\,100$~cm$^2$ & $130\,000$~cm$^2$ \\
        Wavelength range & $100-300$~nm & $100-300$~nm & $100-300$~nm & $100-300$~nm \\
        Spectral resolving power & $R \sim 114\,000$ & $R > 15\,000$ & $R > 45\,000$ & $R > 45\,000$ \\
        Detector dynamic range & 136~cts\,s$^{-1}$\,px$^{-1}$ & 500~cts\,s$^{-1}$\,px$^{-1}$ & 1\,000~cts\,s$^{-1}$\,px$^{-1}$ & 10\,000~cts\,s$^{-1}$\,px$^{-1}$ \\
        \noalign{\smallskip}
        \hline
    \end{tabular}
\end{table*}




\newpage
{\bf Acknowledgements.} L.A.D.S. acknowledges the often-overlooked labor of the custodial, facilities, information technology and security staff at STScI -- this research would not be possible without them. This research is based on observations made with the NASA/ESA Hubble Space Telescope. The data are openly available in the Mikulski Archive for Space Telescopes (MAST), which is maintained by the Space Telescope Science Institute (STScI). STScI is operated by the Association of Universities for Research in Astronomy, Inc. under NASA contract NAS 5-26555. This research made use of the NASA Exoplanet Archive, which is operated by the California Institute of Technology, under contract with the National Aeronautics and Space Administration under the Exoplanet Exploration Program. R.E. carried out research at the Jet Propulsion Laboratory, California Institute of Technology, under a contract with the National Aeronautics and Space Administration (80NM0018D0004).

\bibliography{author.bib}

\end{document}